\documentstyle[aps,preprint]{revtex}
\begin{document}
\title{Anisotropic Quantum Transport in Layered High--$T_c$ Cuprates}
\author{
 S.N. Evangelou$^{1,3}$, Shi-Jie Xiong$^2$, Ding-Yu Xing$^2$, and
 E.N. Economou$^3$ }
\address{$^1$Department of Physics, University of
Ioannina, Ioannina 451 10, Greece \\ 
$^2$Department of Physics and National Laboratory of
Solid State Microstructures, \\
Nanjing University, Nanjing 210093, China \\ 
$^3$Foundation for Research and Technology, 
 Institute for Electronic Structure and Lasers,
 Heraklion, P.O. Box 1527, 71110 Heraklion, Crete, Greece }  
\maketitle
\begin{abstract}
A random lattice model with dilute interlayer bonds
of density $p$ is proposed to describe the  underdoped 
high--$T_c$ cuprates.  We show analytically via an appropriate 
perturbation expansion and verify independently by numerical 
scaling of the conductance that for any finite $p$ the states remain 
extended in all directions, despite the presence of interlayer
disorder. However, the obtained electronic transport is highly anisotropic
with violent conductance fluctuations occuring in the layering 
direction, which can be responsible for the experimentally 
observed metallic ``in-plane" and semiconducting  
``out-of-plane" resistivity of the cuprates.  
\end{abstract}
\vspace{1cm}


\newpage

In recent years there is a great interest in the problem of Anderson 
localization for anisotropic systems \cite{1,2,3,4,5,6}, 
in connection to the unusual normal--state transport
properties of the layered high--$T_c$ cuprates. 
The in--plane resistivity for most of these materials
exhibits metallic behavior,  inreasing linearly with 
temperature over a wide temperature range,
while the out--of--plane resistivity is reminiscent
of semiconductors increasing rapidly at low temperatures \cite{7,8,9,10}.
This contrasting behavior of the parallel and the perpendicular
resistivities  was observed in Bi$_2$Sr$_{2-x}$La$_x$CuO$_y$
far below $T_c$, down to the lowest experimental  temperature
\cite{11}. Logarithmic divergencies of the corresponding
resistivities accompanied by a nearly constant anisotropy ratio are,
instead, observed in the underdoped La$_{2-x}$Sr$_x$CuO$_4$
suggesting an unusual  three-dimensional ($3D$) insulator.

There is currently no consensus on the explanation
of the peculiar $ab$-plane ``metallic"  and the $c$--axis
``semiconducting" transport properties of the cuprates,
although it is commonly believed that the anisotropic conductivity
is, somehow, related to their layered  structure. It is also widely
thought that understanding the peculiar $c$--axis transport
might have important consequences for the high--$T_c$
superconductivity in the cuprates.  
In this respect, many theoretical models 
investigate localization in the 
presence of structural anisotropy, for example,
the  Anderson model with diagonal disorder
and anisotropic hopping $t$ between the $2D$ planes was recently shown
to display very different localization lengths in the parallel
and the perpendicular directions but the same 
critical behavior \cite{6}.
Although this is in agreement with the scaling theory of localization
for anisotropic systems\cite{12,13,14,15} is rather inconsistent
with the observed contrasting resistivities of the cuprates, 
which has often been regarded as evidence for their non--Fermi liquid 
nature \cite{16}.

The intense study of the cuprates did not remove many conflicting 
discrepancies between theory and experiment so that 
a proper description of anisotropic disorder is strongly desirable.
It must be pointed out that previous  
disordered site potentials with fixed but different
hoppings in the lattice axes
imply anisotropy only  in the band structure
and not in the disorder configurations which remain isotropic.
We propose that in order to understand the high-$T_c$ cuprates 
disorder in the anisotropy must be also
incorporated. Anisotropic site randomness 
in a form resembling a  random superlattice 
with lateral inhomogeneities gave anisotropic 
localization if the anisotropy was below a critical value,
even for arbitrarily small disorder \onlinecite{5}.
However, this approach is not a very suitable description for the 
cuprates either, since the CuO$_2$ planes
are believed to be identical with no superlattice--like disorder 
in the layering direction. The  important contribution
to the $c$-axis 
transport in these materials is expected to arise from electron
scattering in the ``insulating'' layer
between the conducting CuO$_2$ layers\cite{7,8,9,10}. 
On the other hand, in almost all high-$T_c$ cuprates   
doping impurities or oxygen vacancies 
always occupy the insulating layers between the 
conducting ``pure" CuO$_2$ planes which leads to  
interlayer disorder. 

Having the above considerations in mind
we propose a very simple $3D$
model with anisotropy both in the band structure 
and the disorder configuration. The system consists of 
perfect parallel lattice planes randomly 
connected by interplane bonds described by the Hamiltonian 
\begin{equation}
\label{ham}
H=\sum_{l}\sum_{\langle {\bf m},{\bf n} \rangle }
|l,{\bf m}\rangle \langle l,{\bf n}|
   + \sum_{l,{\bf m}} t \rho_{l,{\bf m}}(|l,{\bf m}
\rangle \langle l+1,{\bf m}|+
   \text{H. c.}),
\end{equation}
where  ${\bf m}$, ${\bf n}$ are the site indices in every plane and 
$l$ the plane index.     
The first term in Eq. (1) describes nearest-neighbor intraplane hopping 
of unit strength which  sets  the energy unit. 
The second term denotes binary alloy--type of disorder 
which amounts to  interplane hoppings $t$ placed 
at the random planar positions ${\bf m}$, 
implying $100(1-p)\%$ broken longitudinal bonds which inhibit particle 
migration in the layering direction.  The random variable $\rho_{l,{\bf m}}$ 
obeys the distribution 
\[
P(\rho)=(1-p) \delta(\rho) + p \delta(1-\rho),
\]
where $p$ is the perpendicular bond density. 
As a first step towards the explanation 
of the normal--state properties of these materials 
diagonal disorder is ignored so that the in--plane lattices are assumed 
perfect and the disorder represented by $p$ is 
due to impurities or oxygen vacancies 
in the insulating layer among the $2D$ planes. 
The electrons propagating in this system encounter
anisotropy in the disorder due to the
random interplane links, in addition to 
band structure anisotropy due to $t$ being different than one.

A convenient representation for $H$ is the Bloch-Wannier basis 
\begin{equation}
|{\bf k}_{\parallel},l\rangle = \frac{1}{\sqrt{N_{\parallel}}}
\sum_{\bf m}e^{i{\bf k}_{\parallel}\cdot {\bf m}}|l,{\bf m}\rangle ,
\end{equation}
where ${\bf k}_{\parallel}$ is the parallel momentum 
which characterizes the periodic plane  states and 
$N_{\parallel}$ the number of sites in every plane. 
In order to investigate the possibility of localization  
across the planes we define the Green function 
\begin{equation}
G_{\parallel}({\bf k}_{\parallel},l;
{\bf k}'_{\parallel},l';t)\equiv -i\theta (t)
 \langle [ c_{{\bf k}_{\parallel},l}(t),c^{\dagger }_{{\bf k}'_{\parallel},
 l'}(0)]_+\rangle ,
\end{equation}
whose  diagonal elements give the probability
for finding an electron on layer $l$  
at time $t$ with  momentum ${\bf k}_{\parallel}$, 
if initially ($t=0$) was on the same layer with the same momentum. 
The Fourier transform of the diagonal Green function reads
\begin{equation}
G_{\parallel}({\bf k}_{\parallel},l;{\bf k}_{\parallel},l;E)=
\frac{1}{E-\Sigma_{\parallel} ({\bf k}_{\parallel},l,E)}
\end{equation}
with the self-energy $\Sigma_{\parallel}$  written as \cite{17,18}
\begin{equation}
\label{sig}
\Sigma_{\parallel} ({\bf k}_{\parallel},l,E)=
\epsilon_{\parallel} ({\bf k}_{\parallel}) +
\sum_{n=1}^{\infty }\sum_{j}T_j^{(n)},
\end{equation}
where the $n$-th order term is a sum over all paths $j$ starting and 
ending on the initial layer state with
\begin{equation}
T_j^{(n)}=V({\bf k}_{\parallel},l;{\bf k}_{\parallel_1},l_1)
\prod_{i=1}^{n}\frac{V({\bf k}_{\parallel_i},l_i;{\bf k}_{\parallel
_{i+1}},l_{i+1})}{E-\epsilon_{\parallel} ({\bf k}_{\parallel_i})}
\end{equation}
and ${\bf k}_{\parallel_{n+1}}={\bf k}_{\parallel},  l_{n+1}=l$. 
The perfect plane Bloch states characterized by  
${\bf k}_{\parallel_i}$ have energy  
\begin{equation}
\epsilon_{\parallel} ({\bf k}_{\parallel_i}) = 2\cos (k_{x_i}) +2\cos(k_{y_i}), 
\end{equation}
while the interplane hopping which appears in Eq. (6)
in the plane-diagonal representation takes the form 
\begin{equation}
V({\bf k}_{\parallel_i},l_i;{\bf k}_{\parallel_{i+1}},l_{i+1}) 
=\frac{t}{N_{\parallel}}\sum_{\bf m} e^{i({\bf k}_{\parallel_i}-
{\bf k}_{\parallel_{i+1}})\cdot {\bf m} } \delta_{l_{i+1},l_i \pm 1},
\end{equation}
with the sum over ${\bf m}$ denoting only the layer sites having    
bonds present to the nearest neighbor layers.   
It is seen from Eq. (8) that the diagonal in momentum
(${\bf k}_{\parallel_i}= {\bf k}_{\parallel_{i+1}}$) 
matrix elements are  exactly  $pt$  and  
for a given order $n$ if $E$ lies within the $2D$ band  
$\epsilon_{\parallel}({\bf k}_{\parallel_i})$ 
the most divergent term of Eq. (5) comes
from the path which has intermediate states 
$|{\bf k}_{\parallel_i}, l_i\rangle$ 
with $\epsilon_{\parallel}({\bf k}_{\parallel_i}) = E$. The  
corresponding term is $(pt/(E-\epsilon_{\parallel}({\bf k}_{\parallel_i}))^{n}$
and nearest neighbor plane states with the  
same ${\bf k}_{\parallel}$ are predominantly connected. 
It can be also seen that momentum scattering is always accompanied by 
interlayer hopping since  a change of ${\bf k}_{\parallel}$
always leads to a change of layer.  
Therefore, $H$ can be tranformed into an  
``undisturbed" term  $H_0$ which 
represents a perfect anisotropic $3D$ lattice 
with on-plane (interlayer) hopping $1$ ($pt$), plus  
the rest which forms a ``random'' $H_1$ hamiltonian part
with matrix elements  
\begin{equation}
\label{h1ele}
v({\bf k},{\bf k}')=\left\{ 
\begin{array}{l}
\frac{t(e^{ik_z}+e^{-ik'_z})}{N}\sum_l\sum_{{\bf m}_l}
e^{i({\bf k}_{\parallel}-{\bf k}_{\parallel}')\cdot {\bf m}_l +i(k_z-k_z')l}, 
\text{ for } {\bf k}_{\parallel }\neq {\bf k}_{\parallel }', \\ 
0, \text{ for } {\bf k}_{\parallel } = {\bf k}_{\parallel }',
\end{array} \right. 
\end{equation}
expressed in the $H_0$-diagonal basis 
$|{\bf k}\rangle = |{\bf k}_{\parallel},k_{z}\rangle$, where
$N$ is the total number of lattice sites, $l$ the layer index, 
$k_z$ the perpendicular momentum  and  the  
sum over ${\bf m}_l$ runs over sites which are 
connected to nearest neighbor planes by  
interplane bonds.   In the adopted basis $H_1$ has only off-diagonal
matrix elements in the momentum ${\bf k}_{\parallel}$
between planes since the diagonal elements
were already included in $H_0$.

In order to determine the properties   
for the eigenstates of $H_0$ characterized by ${\bf k}$, with
corresponding eigenvalues $\epsilon ({\bf k})$, we  
can expand the corresponding self-energy 
$\Sigma ({\bf k}, E)$ as in Eq. (\ref{sig})
via the expansion parameter 
\begin{equation}
\label{exp}
T(E,{\bf k}')=\sum_{{\bf k}''}\frac{v({\bf k}',{\bf k}'')}{E-\epsilon 
({\bf k}'')} 
\end{equation}
and if Eq. (9) is substituded into  Eq. (10) 
\begin{equation}
\label{exp2}
T(E,{\bf k}')=\sum_{{\bf k}''}\sum_l\sum_{{\bf m}_l}\frac{t(e^{ik'_z}+e^{-ik''_z})
e^{i({\bf k}_{\parallel}'-{\bf k}_{\parallel}'')\cdot {\bf m}_l +i(k_z'-k_z'')l}}
{N[E-\epsilon ({\bf k}'')]}
\end{equation}
evidently vanishes on average taken over a large number 
of configurations, since all the on-plane site positions will be 
exhausted for the present interplane bonds by the additional
sum for the average which  leads to cancellations in Eq. (\ref{exp2})
from Bloch theorem. 
A typical $T$ for a given random configuration   
fluctuates around zero with a variance $\sigma^2_{\text{sum}} 
\sim Q\sigma ^2$, where $\sigma^2 $ is the variance of a single term 
and $Q$ the total number of terms in the sum. 
According to the denominator $E-\epsilon ({\bf k}'')$ 
the most important contribution  
comes from states with momenta on an equal-energy surface in the 
${\bf k}$-space whose number is of the order of 
$N_{\parallel}=L^2$. 
The total number of present interplane bonds is $pL^3$  
which gives  $Q\sim pL^5$, so that for a given configuration 
the sum is $\sim \sqrt{p}L^{\frac{5}{2}}$. Finally, 
by including the normalization factor $N =L^3$ in the denominator 
$T(E,{\bf k}') \sim \sqrt{p}L^{-\frac{1}{2}}$ 
which guarantees rapid convergence of the self-energy in 
${\bf k}$-space  to 
\begin{equation}
\Sigma ({\bf k},E) \simeq \epsilon ({\bf k}) +
{\sum_{\bf k'}} {\frac {|v({\bf k, k'})|^{2}} 
{E-\epsilon ({\bf k}') }}  .
\end{equation}
The configuration average 
$ \langle Im {\sum_{\bf k'}} {\frac  {|v({\bf k, k'})|^{2}} 
{E-\epsilon ({\bf k}') -i 0^{+}}} \rangle$ 
is also computed as a function of $p$ and the obtained 
semicircular form can be fitted 
by  $\rho(E) t^{2} p(1-p)$, with $E$  within the 
$H_0$ band with $\rho (E)$ the density of states. 
This allows to estimate the 
lifetime of states  $\tau \sim \frac{1}{\rho (E)t^{2}p(1-p)}$ 
and the corresponding  mean free paths 
\begin{equation}
\lambda_{\parallel} = {\frac {\tau u_{\parallel}^{2}}{u}} \simeq
\frac{1} {\rho (E)t^{2}p(1-p)\sqrt{2+p^{2}t^{2}}},
\end{equation}
\begin{equation}
\lambda_{\perp} = {\frac {\tau u_{\perp}^{2}}{u}} \simeq
\frac{p} {\rho (E)(1-p)\sqrt{2+p^{2}t^{2}}},
\end{equation}
by taking into account the Fermi velocites $u_{\parallel (\perp)}$.
It can be seen that for small-$p$ the parallel (perpendicular) 
mean free path is proportional to $1/p$ ($p$) so that  
the interlayer disorder affects more the perpendicular transport
although cannot localize the states even in this direction. 

The predicted extended nature of the wave functions is 
independently verified by numerical scaling of the conductance 
in finite  systems. The 
parallel (perpendicular) conductance  $G_{\parallel (\perp )}(L)$ 
as a function of the linear size $L$ is obtained for a 
cubic system at the Fermi energy $E$ 
from the multichannel Landauer--Buttiker formula \cite{19,20,21}
\begin{equation}
G_{\parallel (\perp )}(L) = (e^2/h)\text{Tr}(T^+_{\parallel 
(\perp )}T
_{\parallel (\perp )}),
\end{equation}
where $T_{\parallel (\perp )}$ is the  
transmission matrix for electronic
propagation along the parallel (perpendicular) direction 
which can be obtained  by transfer matrix 
techniques.  The matrix $T_{\perp}$ turns out to be 
singular due to the presence of zeros for the missing bonds.
This difficulty can be dealt with succesfully
if we introduce hard-wall boundary conditions in the 
parallel directions and assign a unit incident wave amplitude
for one channel, with  zero for the rest, at a time.  
The recursion relations for the corresponding  
wave function coefficients are easily established  
along the parallel direction while along the difficult perpendicular 
direction the assignment made for the 
incident waves is used as a 
boundary condition. This is a convenient method which allows to consider
propagation in the difficult layering direction by a transfer matrix
product in the easy parallel direction since    
the $L^{2}$ equations  for the the unknown coefficients 
can be solved.
The absence of missing bonds in the parallel direction makes 
the recursion relations no longer
singular and $T_{\perp}$ can be obtained for the
output channels, for every assignement made 
for the input channels.   In the calculations we have taken
averages over up to a $5000$ random configurations 
in each case to suppress fluctuations.

In Figure 1 we show the scaling behavior  of the
dimensionless conductance $g_{\parallel (\perp )}(L) = G
_{\parallel (\perp )}(L)/(e^2/h)$ with  
anisotropic interplane coupling $t=0.3$ for
various bond densities $p$. The parallel
conductance increases ballistically
($\sim L^{2}$) for small $L$ and linearly for higher $L$. 
In the large size ($L\to \infty$) limit  the scaling function
$\beta (L) = d\ln g/d\ln L$ 
is positive for $g_{\parallel (\perp)}$,
which implies extended states in both  directions 
for any  $p$ in agreement  with our analytical results.
However, the transport behavior found is essentially different 
in the two directions. 
In Fig. 2 we show the energy dependence of the conductance 
for  $p=0.5$ and $t=0.3$ where 
the ratio of the conductances is close to the estimate   
$g_{\parallel }/g_{\perp }\simeq (t_{\parallel}/t_{\perp})^2$
\cite{6} for $t_{\parallel}=1$ and $t_{\perp}=pt=0.15$.  
Another key  finding  from Fig. 2 is a rather smooth
$g_{\parallel}(E)$  while $g_{\perp}(E)$ displays violent 
oscillations as a function of $E$. 
The dips in $g_{\perp}(E)$ can be ragarded as ``minigaps" 
which might have an  effect similar to a semiconductor 
leading to  insulating kind of behavior for the 
out-of-plane conductivity when the Fermmi energy is varied.   
 
We argue that the proposed model is closely related to 
high--$T_c$ cuprates. As the temperature increases the 
inelastic scattering in these materials due to phonons, spin waves or other
excitations within the CuO$_2$ layers 
can cause a decrease of the
characteristic inelastic scattering lengths $l_{in}$. If
the temperature is so high that $l_{in}$ becomes smaller
than the mean free path caused from the ``random'' 
Hamiltonian $H_1$ the transverse 
conductivity becomes perfectly metallic.
Experiments for  Bi$_2$Sr$_2$CaCu$_2$O$_8$ 
and underdoped La$_{2-x}$Sr$_x$CuO$_4$, YBa$_2$Cu$_3$O$_{6+x}$ 
give  out--of--plane resistivity which has a  semiconductor--like
temperature dependence at low temperatures  and a linear--in--$T$ behavior 
at high temperatures with  characteristic
crossover temperature $T^{*}$ between the two regimes,
which decreases by increasing doping in La$_{2-x}$Sr$_x$CuO$_4$
and YBa$_2$Cu$_3$O$_{6+x}$ \cite{7,8,9,10}. 
We notice that if we relate the 
bond density $p$ with the doping density 
of the high-$T_c$ materials  the obtained $p$-dependence of the 
perpendicular mean free path can be used to 
explain qualitatively  the  reported experimental behavior.
It must be pointed out
that the relation between the cuprate doping density  and
the bond density $p$ is very natural, since an increase in the
number of the doping impurities 
or the oxygen atoms in the layer between two CuO$_2$ planes 
increases the number of hopping paths between the two planes.
As $T^{*}$ decreases further below the
superconducting transition temperature $T_c$ the out--of--plane 
normal--state resistivity also becomes metallic,
which has recently been observed in high--quality
single crystals of YBA$_2$Cu$_3$O$_7$ and other
high--$T_c$ cuprates \cite{22} corresponding to the absence of
disorder ($p\approx 1$ in the proposed model) with almost
infinite perpendicular mean free path. This effect can be 
observed only in the perpendicular direction since the parallel
mean free path is always much longer and the inelasting scattering more 
dominant in ths direction.

In summary, we have introduced a simple layered lattice model
for high-$T_c$ cuprates based on the fact that  these materials 
consist of pure CuO$_2$ planes while doping 
impurities or oxygen vacancies occupy randomly 
the interplane hopping paths.  We introduce 
anisotropic disorder  described by 
the interplane bond density $p$, in addition to the usual
anisotropic band structure expressed 
via the interplane hopping strength $t$.  
In the absence of diagonal disorder we show  extended states in both  
directions but the obtained 
mean free path and the conductance in the perpendicular 
direction is much smaller than the 
parallel one. Moreover, the perpendicular conductance  
fluctuates strongly  as a function of  
energy, which might lead to an insulator-like 
temperature dependence of the conductivity in this direction. 
To conclude, we have
demonstrated anisotropic transport which can account 
for experimental facts of high-T$_c$ 
cuprates. More work is needed in order to
understand the corresponding metal-insulator transition 
of layering materials also in the presence of diagonal disorder
and/or a homogenous magnetic field.


{\bf Acknowledgments} 
 
 This work was supported in part  by a TMR grant  and
also a $\Pi$ENE$\Delta$ Research Grant of the Greek Secretariat
of Science and Technology. 

 {}

\newpage\clearpage
\begin{figure}[h]

{\bf FIG.1}. The $g_{\parallel}$ as a function of the linear
system size $L$ for a cubic system of parallel planes 
with randomly placed interplane 
bonds of density $p$ and strength $t=0.3$.   
In the inset $g_{\perp}$ is shown for the same system.  
 
{\bf FIG.2}. ${\bf (a)}$ The energy-dependent $g_{\parallel}$  
for a layered system with $L=10,15$ and $t=0.3$, $p=0.5$.
${\bf (b)}$ A much smaller $g_{\perp}$ is shown
which exhibits violent oscillations as a function
of energy. 

\end{figure}
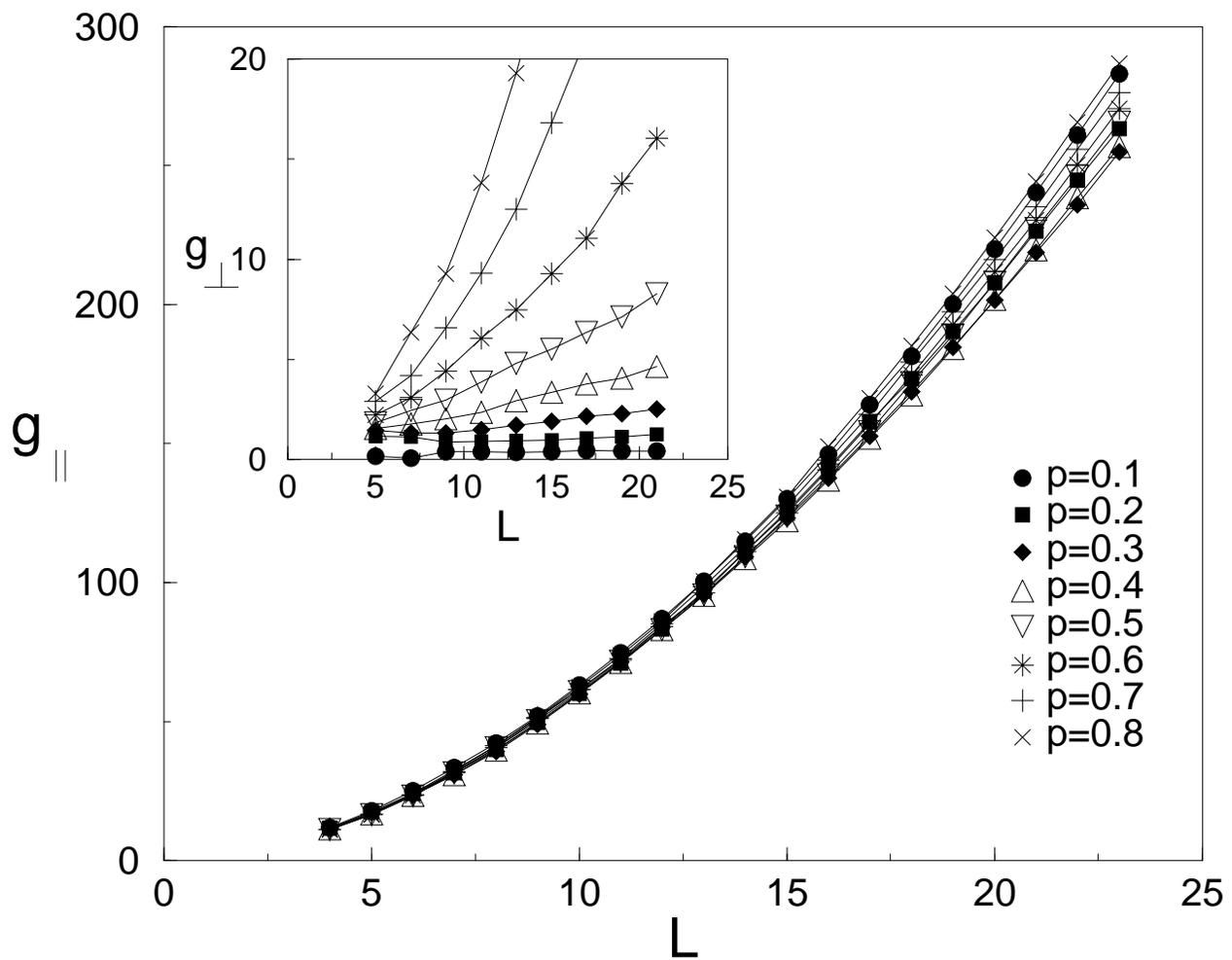
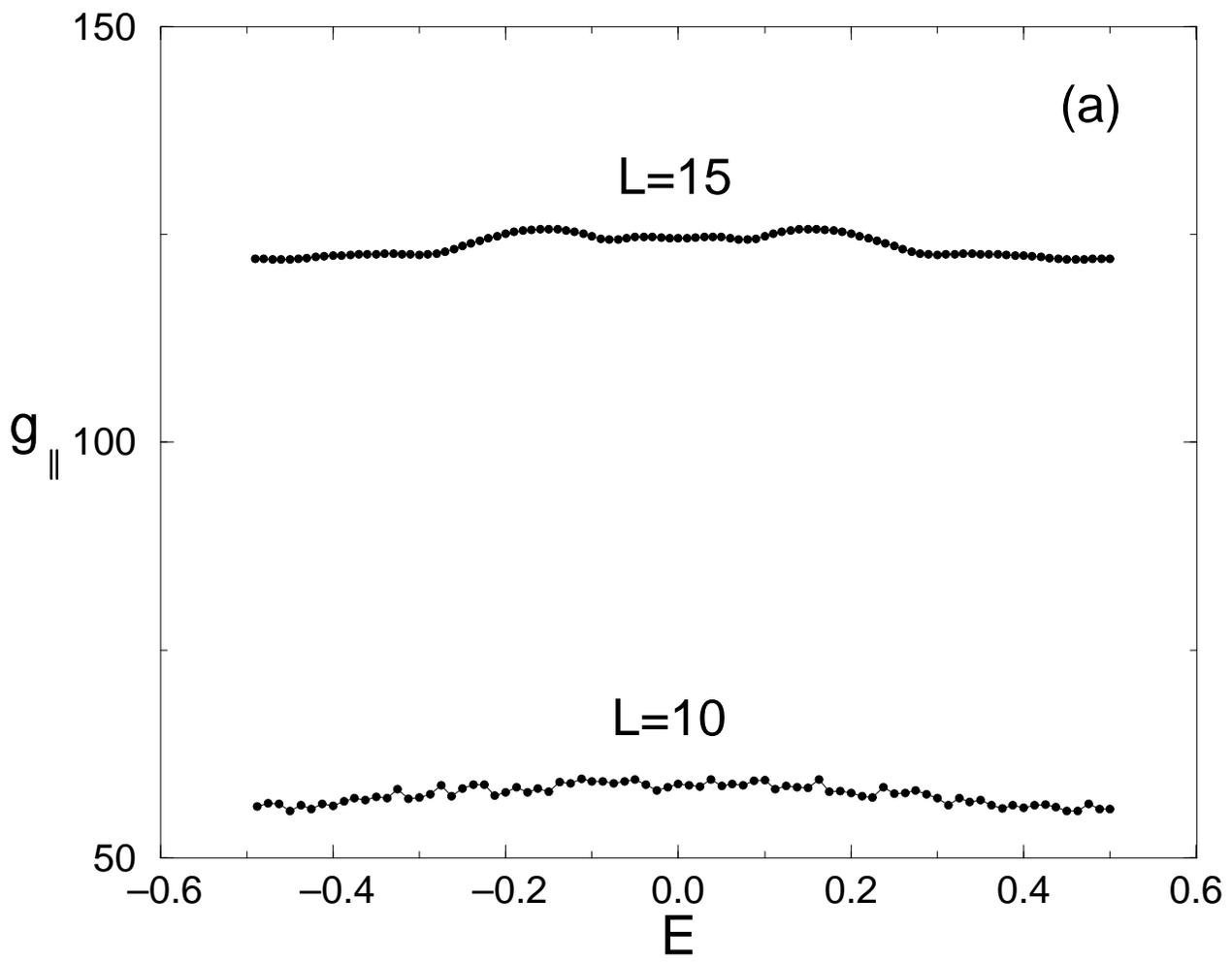
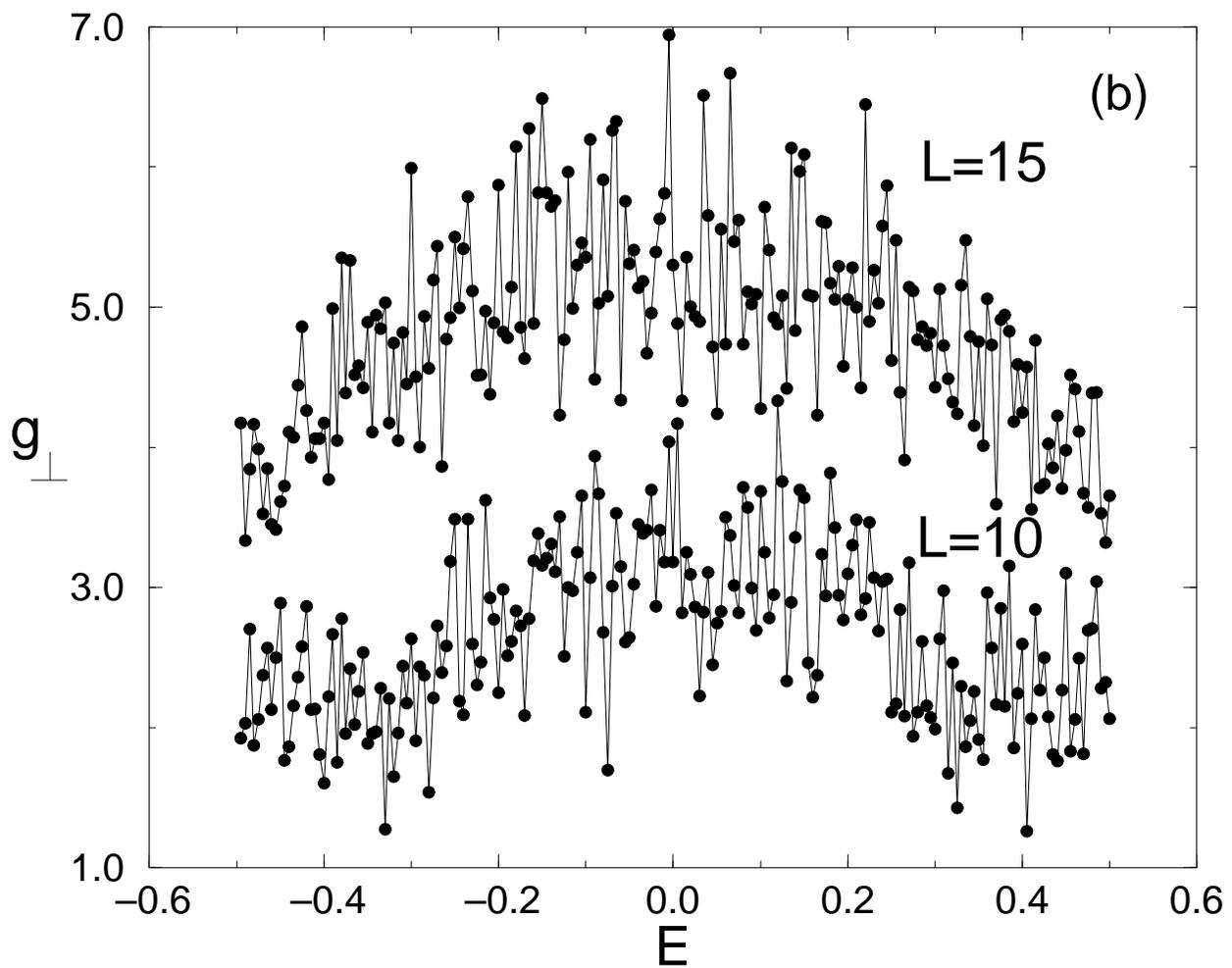
\end{document}